\documentclass[number,12pt,english]{elsarticle} 
\usepackage[T1]{fontenc}
\usepackage[utf8]{inputenc}
\usepackage{geometry}
\usepackage{amsmath}
\usepackage{graphicx}
\usepackage{esint}
\usepackage{hyperref}
\usepackage{graphicx}
\usepackage{array}
\usepackage{amssymb}
\usepackage{multirow}
\usepackage{adjustbox}
\usepackage{longtable}
\usepackage{varwidth}
\usepackage{textcomp}
\usepackage[active]{srcltx}
\usepackage{babel}
\usepackage{array}
\usepackage{multirow}
\usepackage{varwidth}
\usepackage{gensymb}

\usepackage{setspace}
\makeatletter

\providecommand{\tabularnewline}{\\}

\addto\extrasfrench{%
   \providecommand{\fg}{\ifdim\lastskip>\z@\unskip\fi~\frqq}%
}

\makeatother

\usepackage{setspace}
\journal{Earth, Planets and Space}

\begin{document}

\begin{frontmatter}{}

\title{Simulation of laser travel-time on Mercury for BELA}

\author[label1]{J. Barron \corref{cor1}}
\ead{jean.barron@universite-paris-saclay.fr}

\author[label1,label2]{F. Schmidt}
\author[label1]{F. Andrieu}

\author[label3,label4]{G. Nishiyama}
\author[label3]{A. Stark}
\author[label3]{H. Hussmann}

\address[label1]{Universit{\'e} Paris-Saclay, CNRS, GEOPS, 91405, Orsay, France}
\address[label2]{Institut Universitaire de France (IUF)}
\address[label3]{Institute of Space Research, Deutsches Zentrum für Luft und Raumfahrt, Rutherfordstr. 2, 12489 Berlin, Germany}
\address[label4]{Department of Earth and Planetary Science, Graduate School of Science, The University of Tokyo, Tokyo, 113-0033, Japan}

\cortext[cor1]{Corresponding author}

\begin{abstract}
Recent laser altimeters are able to not only measure the ranging distance between the spacecraft and the surface but also the full time-of-flight of the photons or pulse shape. This new capabilities allows to measure the intra-footprint properties: surface slope distribution and surface microtexture. Here we simulate and discuss for the first time the effect of surface microtexture, especially for ice covered surface with longer penetration depth. Using the WARPE simulation software, two kind of microtextures are simulated: compact slab and granular. Laser pulse shape for an ideal instrument is simulated using physical properties such as the grain size, material composition, thickness, compacity (filling factor, porosity) rather than radiative properties. The effects of these parameters on the pulse shape are discussed as well in the range that could be possibly be observed with actual BELA measurement. Finally, examples of WARPE's simulated pulse shapes are used as input in the precise simulation chain of the BELA measurement output, to further assess the capability to detect variation in surface microtexture.

\end{abstract}


\end{frontmatter}
\newpage
\section{Introduction}

The BepiColombo mission \cite{Benkhoff_2021} carries on board the BepiColombo Laser Altimeter (BELA) \cite{Thomas_2021} with one of the scientific objectives being "the surface roughness, local slopes and albedo variations globally and within permanently shaded craters near poles". The permanently shadowed region have previously been studied on Mercury with the first full-disk radar mapping presented and discussed in \cite{Slade1992} highlighting that polar regions present highly reflective material. These reflectors were later identified inside craters at both north pole \cite{Harmon2011} and south pole \cite{Chabot_2018} of Mercury. PSRs present stable cold regions according to thermal models based on topographic data \cite[e.g.,]{Paige2013, Gl_ser_2022} where the maximum temperature is estimated around 100K even though some crater at lower latitude can reach higher temperature up to 210K \cite{Cambianica2024}. The age of radar-bright region seems younger than region without radar-bright material and crater ejecta for PSR crater \cite{Bertoli_2024,Bertoli_2025}. Thus the high latitude PSRs present strong argument for possible stable water ice inside those craters. In situ properties remain unknown. \cite{Susorney_2019} proposed a maximum thickness of 15 meter for PSR-ice. Assumptions for the PSR non-icy terrains can be made using the global surface albedo of Mercury from Mariner 10 \cite{DENEVI_2008} and MESSENGER \cite{Domingue_2011} and also approaching the albedo of underlying material beneath a hypothetical icy terrain. To compare with a similar object, the size of scattering grains on the moon have been estimated in \cite{mckay1991lunar} and \cite{Szabo_2022} between 60 and 80 microns, with a large forward scattering \citep{Warell_2004}. Regolith porosity has also been discussed to be around 85\% \citep{Szabo_2022}. Another major question is regarding the composition of the ice. Although water ice is hypothesized to be the predominant species in PSR since they act as cold traps \cite{Rubanenko_2018, Rubanenko_2019}, pure water ice seems geologically unrealistic as \cite{Barker_2022} used MLA albedo measurements in Prokofiev's PSR to suggest the presence of darker (maybe organic) components. 
The composition is unknown but water ice remains the most probable candidate from the surface temperature \cite{Paige2013}. Impurities may be also present \cite{Glantzberg_2023}. While coronene is a suitable candidate \cite{Hamill_2020} among other candidates like aromatic hydrocarbons \cite{Zhang_2009,Zhang_2010}, the true nature of volatiles in PSR remains unknown.
The origin of PSR volatiles component is also an active scientific debate. Mainly, two scenarios are possible: internal volatile (for instance from volcanic degassing) \cite{Crotts_2009,Lawrence2017} or external (from solar wind interactions or comets and asteroids) \cite{Lucey_2009,Lawrence2017}. In both case, volatiles seems cold-trapped in the PSR due to the very low temperature. Identifying these impurities and characterizing the texture of the PSR could lead to decipher their source. 
Another interesting geological feature potentially linked to volatiles is hollows where various species have been identified such as chloride \cite{Vaughan2012}, sulfides like MgS and CaS \cite{Barraud_2020, Barraud_2023} or even graphite \cite{Murchie_2015, Emran_2025}. 

Previous study \cite{Doute_reflectancemodel_JGR1998, Pilorget2013, Andrieu_2016} demonstrated that microtextures and physical properties influence the reflectance received by an instrument. Pulse shape being the power transmitted and received as a function of time \cite{Bufton1989}, the same influence with radiative properties can be observed on pulse shape \cite{Barron_2025}. The idea here is to unravel geological constrain to better characterize these structures using supplementary informations given by the received pulse shape.

This paper focuses on describing the influence of medium's microtexture and physical properties (composition, impurities, porosity, grain size, compact/granular...) on the waveform expected on BELA, related with possible volatiles presence at Mercury's surface. We used the WARPE \cite{Barron_2025} model to simulate synthetic waveforms and then focus on the most favorable conditions that could be observed by BELA. The scientific target is PSR but this work could also be potentially applicable to other material in Mercury or other ices in the solar system.

\section{Method}

\subsection{Simulation}
\label{subsec:Simulation}

For this study we use WARPE (Waveform Analysis and Ray Profiling for Exploration) model \cite{Barron_2025, WARPE}: a Monte Carlo ray-tracing model that computes the reflectance as a function of the travel-time for granular and slab media configuration. The result of the simulation gives an equivalent of the return pulse shape in a waveform laser signal. The model uses radiative parameter such as the optical depth $\tau$ or the single scattering albedo $\omega$ describing the optical behavior of a medium and the optical constant ($n$,$k$) for the Fresnel reflection.
We used here the parametrization of the radiative parameters from physical parameters from \cite{Hapke1993}, \cite{Andrieu2015a} and \cite{doute1998}. Thanks to these equations, the soil's in-situ microtexture (grain size, porosity...) is converted into radiative transfer parameters ($\tau$, $\omega$) used as input in WARPE. Hereafter the main equations summarized. The full list of physical parameters is shown in Table \ref{tab:physical_param}.

\begin{table}
\centering
\centerline{\hfill{}%
\begin{tabular}{|c|c|}
\hline 
Symbol & Description\tabularnewline
\hline 
$h$ & thickness\tabularnewline
\hline 
$d$ & grain size (grain diameter)\tabularnewline
\hline 
$\gamma_{c}$ & compacity or filling factor\tabularnewline
\hline 
$\rho_{v}$ & volumetric proportion\tabularnewline
\hline 
$n$ & real component of optical constant\tabularnewline
\hline 
$k$ & imaginary component of optical constant\tabularnewline
\hline 
$\lambda$ & wavelength of the light interacting with the medium\tabularnewline
\hline 
\end{tabular}\hfill{}}

\caption{Physical parameter used in the computation of $\tau$ and $\omega$ used then as entry parameters of WARPE}\label{tab:physical_param}
\end{table}

The optical depth $\tau$ of a granular layer with inclusion is: 

\begin{equation}
\tau=-\mathcal{N}\frac{\ln(1-\gamma_{c})}{\gamma_{c}}<\sigma_{e}>\upsilon
\end{equation}

The optical depth $\tau$ of an ice slab with inclusion is: 

\begin{equation}
\tau=\left(\mathcal{N}<\sigma_{e}>\frac{\ln\gamma_{c}}{\gamma_{c}-1}+a_{m}\right)\upsilon
\end{equation}
where $\mathcal{N}$ is the total number of inclusions per volumetric unit, $<\sigma_{e}>$ is the mean extinction cross section, $a_{m}$ is the absorption coefficient and $\upsilon$ is the physical length of the path, $\gamma_{c}$ is the compacity (also sometimes noted as filling factor). The porosity is thus $1-\gamma_{c}$ and could be filled by another material or by void.

In the WARPE model, the light is described as $N$ number of rays following geometric optic. Grains are considered near-spherical and media are homogeneous (same statistical properties everywhere) but could contain heterogeneities. For compact slab media, Fresnel coefficients using both effective optical constants $n_{eff}$ and $k_{eff}$ are computed for interactions at interface (refraction and reflection). For granular media, there is no defined interface therefore the computation of Fresnel coefficients becomes irrelevant. In both situation the light velocity is controlled by $n_{eff}$. The bottom condition is considered as lambertian with albedo $A$. This assumption is valid if the impurities are present at the surface with a surface proportion similar to the volumetric proportion.

WARPE model estimates the total optical length of each rays, and converts it to physical length using the following equation:

\begin{equation}
D_{\upsilon}=D_{\tau}\frac{h}{\tau_{0}}
\end{equation}
with $D_{\upsilon}$ is the physical length for each ray within the medium, $D_{\tau}$ is the optical length for each ray, $h$ is the physical thickness of the medium and $\tau_{0}$ is the total optical depth of the medium.  
The travel-time $t$ of each rays follows: 

\begin{equation}
t=D_{\upsilon}\frac{n_{eff}}{c}
\end{equation}
where $n_{eff}$ is the effective refractive index of the medium and $c$ is the speed of light in void. The effective refractive index $n_{eff}$ of the medium is determined from $n$ of each end-member, given their proportions. For compact slab, the formula is:

\begin{equation}
n_{eff}=n_{matrix}\times \gamma_{c}  + \frac{\sum_{i=1} n_{i} \times \rho_{v,i}}{ \sum_{i=1} \rho_{v,i}} \times (1-\gamma_{c})
\end{equation}

With $i$ the index of impurities (each of them characterized by index $n_i$  and volumetric proportion of each impurities $\rho_{v,i}$) and $n_{matrix}$ the index of the matrix. The value of the extinction coefficient $k$ of the medium is also computed with the same equation.

For granular material, the formula is:
\begin{equation}
n_{eff}=\frac{\sum_{i=1} n_{i} \times \rho_{v,i}}{ \sum_{i=1} \rho_{v,i}} \times \gamma_{c} + n_{void} \times (1-\gamma_{c})
\end{equation}

The value for the effective extinction coefficient $k_{eff}$ of the medium is also computed with the same equation.

\subsection{Numerical experiments for ideal BELA simulation}
\label{subsec:Numerical-experiments}

The aim of this article is to provide synthetic simulation of the return pulse of BELA. We will thus estimate the reflectance as a function of time, in a unit of ns$^{-1}$sr$^{-1}$. In these simulation, the instrument is considered ideal, all the electronics and instrumental effects are described by \cite{Nishiyama_PulseShapeRoughness_2026}. The end of this article will provide an example of instrumental effects.

\paragraph{Time pulse width and time sampling}
The emitted BELA laser pulse width is 5 ns and 8 ns (in Full Width Half Maximum) depending on temperature condition and operation configurations \cite{Thomas2007,Althaus_2019} and the sampling interval is 12.5 ns. Here we set a numerical experiment to address the time impulse response of the surface to keep it as general as possible. A convolution with the laser pulse width is thus required to get the correct simulated waveform. Since the laser pulse width is much smaller than the sampling time, this effect will not affect our BELA simulation (see section \ref{sec:BELA_measurement}).

Here we only consider the propagation in the surface, assuming different microtexture and composition. The time-of-flight from the spacecraft to the ground (which is a simple time delay), the effect due to the surface roughness and local slope at the scale of footprint and the effect of BELA data receiver chain are ignored. Those effects are discussed in \cite{Nishiyama_PulseShapeRoughness_2026}.

\paragraph{Beam divergence}
The BELA beam divergence is $60\,\mathrm{\mu rad}$ (full-cone, \cite{Thomas_2021}). It means that in the case of a perfectly flat surface perpendicular to the beam, the specular reflection is at time 0 (no transfer in the medium and the time-of-flight spacecraft/ground is ignored here) and the back and forth direct waves (BF) have the same $60\,\mathrm{\mu rad}$ divergence. In this geometry, the effective radius $a$ of the spatial extension of the beam reflected from the surface back to space again is:

\begin{equation}
a=2H\tan\delta
\end{equation}
with $H$ the altitude of the spacecraft and $\delta$ the divergence of beam.

In nominal BepiColombo operation mode, the altitude of the satellite is between $400\,\mathrm{km}$ and $1400\,\mathrm{km}$, meaning an effective radius $a$ of the spatial extension of the beam from the surface back to space again is between $24.8\,\mathrm{m}$ and $84.0\,\mathrm{m}$.

Meanwhile, the considered telescope aperture diameter $A_r$ is $200\,\mathrm{mm}$ \citep{Thomas_2021}. Assuming that the flux is isotropic within the beam, the ratio between all the energy that is specularly sent back from the surface is defined by:
\begin{equation}
r_{aperture/beam} = \frac{\pi (\frac{1}{2}A_r)^2}{\pi a^2}
\end{equation}

BELA can detect a proportion up to between $1.71\times10^{-5}$ and $1.39\times10^{-6}$ of what a perfectly flat surface specularly sent back.

From \cite{Andrieu2015a,Andrieu_2016}, a specular reflection from icy surface for $\sim0.1$\textdegree~mean slope create a specular lobe of a half aperture cone of 0.25° ; respectively 2.5° half aperture for a roughness of $\sim0.5$\textdegree. Here a roughness of $\sim0.1$\textdegree is arbitrarily used to evaluate all specular reflection:
\begin{equation}
a_{spec}=H\tan(0.25\degree)
\end{equation}

\begin{equation}
r_{aperture/beam} = \frac{\pi (\frac{1}{2}A_r)^2}{\pi a_{spec}^2}
\end{equation}

Thus, in this work, instead of setting WARPE with the actual beam divergence, we used the perfect 0° incidence. This allows us to use the previous rationale to very easily change the specular reflectance intensity to any spacecraft altitude and roughness condition. Please note that the effect only concerns the specular reflection and the Back and Forth direct (noted BF direct), i.e.: the photons reflected back from the bottom interface and leaving the medium at the top. 

To estimate the pulse shape reflectance for specular reflection $R$ in ns$^{-1}$sr$^{-1}$ of an ideal instrument, we used the formula:

\begin{equation}
R(t_{spec}) = \frac{N_{spec} \times r_{aperture/beam}}{N\times \Omega \times \Delta t}  
\end{equation}
With $\Omega$ the aperture angular width in sr$^{-1}$, $t_{spec}$ in travel time of the specular (0, or $\frac{2h}{n_{eff}}$ for BF), $N_{spec}$ is the number of rays in WARPE outgoing at emergence 0°. $\Delta t$ is the size of the temporal bin.

The BF diffuse and the scattering component are not sensitive to the incidence angle divergence since there is always a scattering event that changed the direction. In following simulation, an altitude of $500\,\mathrm{m}$ will considered.

\paragraph{Wavelength}
Since BELA is equipped with a laser at 1064 $\mathrm{nm}$ \cite{Thomas2007}, we select optical constants at this wavelength. We used the optical constant of water ice and CO$_2$ ice from \cite{Schmitt_1998} and the one from sulfur, potential candidates for impurities from \cite{Sasson_1985}. The values are refered in Table \ref{tab:endmember}.

\begin{table}
\centering
\centerline{\hfill{}%
\begin{tabular}{|c|c|c|c|}
\hline 
End-member & refractive index & extinction coefficient & Reference\tabularnewline
\hline 
Void & $1$ & $0$ & -\tabularnewline
\hline 
H$_2$O & $1.30042$ & $1.27200\times10^{-6}$ & \cite{Schmitt_1998} \tabularnewline
\hline 
CO$_2$ & $1.4026$ & $8.47\times10^{-10}$ & \cite{Schmitt_1998} \tabularnewline
\hline 
S & $1.92$ & $9.5\times10^{-6}$ & \cite{Sasson_1985}\tabularnewline
\hline 

\end{tabular}\hfill{}}

\caption{Optical constant at 1064 nm of the end-members used in this study}\label{tab:endmember}
\end{table}

\paragraph{Scattering phase function}
The scattering phase function is assumed to follow the Heynyey-Greenstein law \cite{Henyey1941} with 1 parameter. In this study the value of the lobe parameter is set to $g=0.8$ for forward scattering  ($g=0$ for isotropic in this convention, $g=1$ is strongly forward, $g=-1$ is strongly backward) \cite{Barron_2025}.

\paragraph{Numerical Field of View (FOV)}
 The incidence angle is set 0° (nadir). The emergence is close to 0° but WARPE simulates rays leaving the medium in all directions of azimuth (0°-360°) and emergence (0°-90). However, BELA's receiver has a specific field of view (FOV) of 530 $\mu$rad (full cone). A scaled geometry for WARPE is required so simulations correspond to BELA configuration. Unfortunately, a FOV of 530 $\mu$rad cannot be directly used in WARPE because almost no rays will reach this geometry, even for $N\sim$ million of rays. To approach the true laser FOV measurement, the aim here is to find numerically the largest emergence where the BRDF is locally isotropic. This angle is used as a numerical FOV.

\begin{figure}[h!]
\centering
\centerline{\hfill{}\includegraphics[scale=0.5]{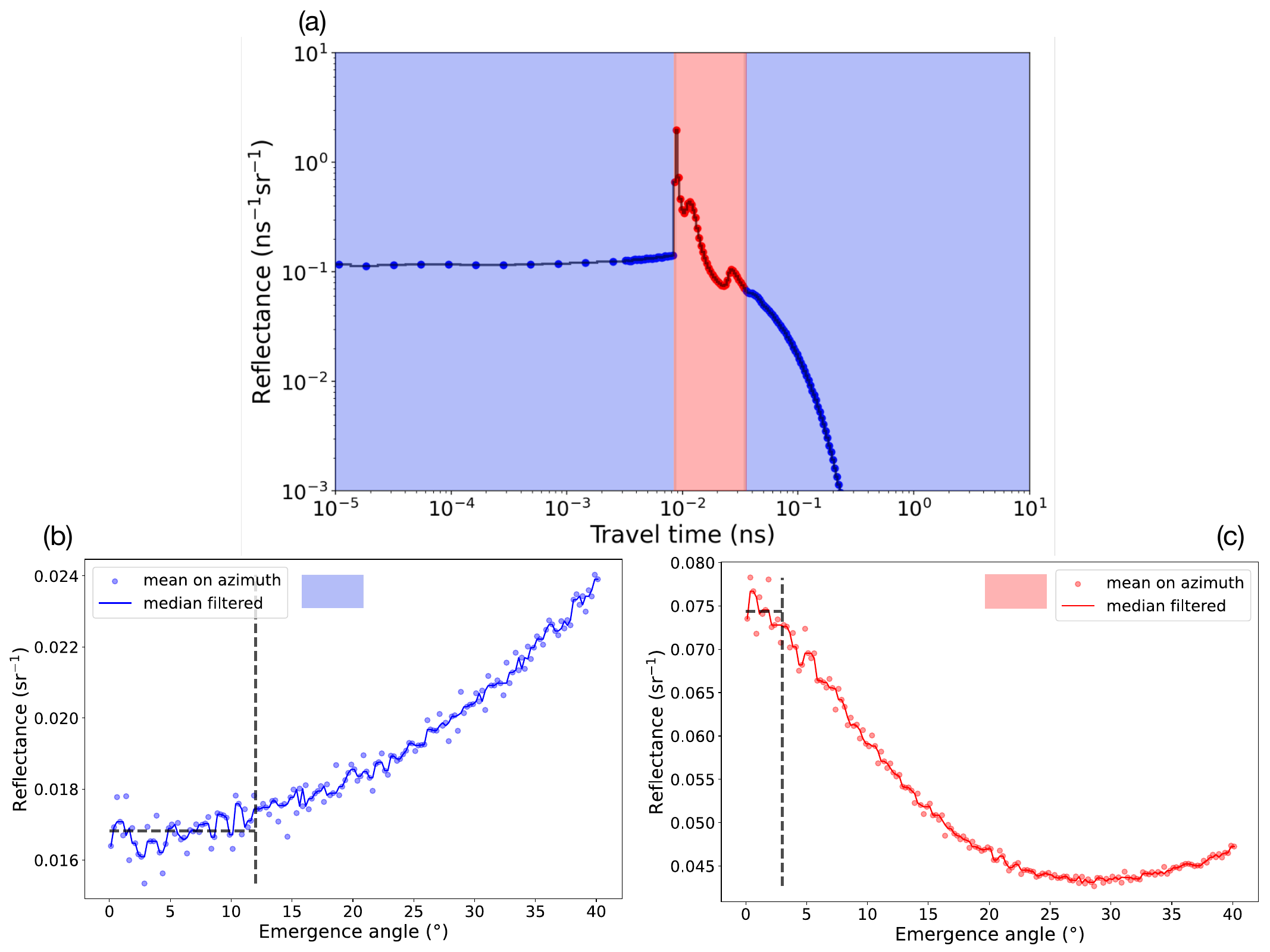}\hfill{}}
\caption{Determination of the numerical field of view. (a) WARPE's simulation for a water ice slab of 1 mm with 1\% of CO$_2$ impurities with  25 $\mathrm{\mu m}$  grain size. Blue domain selects the background scattering ; Red domain selects the back and forth diffuse wave \cite{Barron_2025}. Domains are determined using the characteristics return time for back and forth to properly discriminate its influence on background scattering. Here the specular reflection and back and forth (BF) direct are ignored since these features are geometrically (in perfectly 0° of emergence) and timely condensed. Thus, these graphs only consider rays with at least 1 scattering event. (b) Mean reflectance along azimuth for the background scattering. The reflectance is flat (locally lambertian) from 0° to 12° with a RMS of 0.0006. We set the numerical FOV value at 12°. (c) Mean reflectance along azimuth for back and forth diffuse wave. The reflectance is flat (locally lambertian) from 0° to 3° with an RMS of 0.0023. We set the numerical FOV value at 3°.}\label{fig:FOVnum}
\end{figure}

Figure \ref{fig:FOVnum} shows that fortunately the outgoing rays follow a local lambertian behavior near the zenith direction (0° emergence angle). Since the BF diffuse and the background scattering are not following the same angular dependencies, we choose to define a numerical FOV of 3° and 12° respectively. This angular behavior is not expected to change much with composition. The numerical FOV allow us to mimic the actual behavior of the FOV of $530\,\mathrm{\mu rad}$ with a reflectance unit of ns$^{-1}$sr$^{-1}$ with a much broader FOV.

\paragraph{Media geometry}
For this article, the model assumed an horizontal layer (either compact slab or granular), perfectly perpendicular to the laser beam. A surface slope could be introduced, leading to expected effect: (i) if the slope is larger than the reflected specular beam half opening angle (which is simply $30\,\mathrm{\mu rad}$ the divergence of the BELA laser beam in the case of perfectly flat surface), the laser will not receive any specular reflection anymore but only back and forth diffuse and scattering components (ii) if the latter condition is not fulfilled, the reflectance in ns$^{-1}$sr$^{-1}$ is not changed, assuming that the laser beam is isotropic.

\newpage
\section{Results}
\label{sec:Results}

This section describes various effects of physical properties on the waveform for different compositions as well as different textures: compact slab and granular. These media configurations have been previously described for WARPE  \cite{Barron_2025}. As seen in section \ref{subsec:Numerical-experiments}, the chosen numerical FOV are 12° for the background scattering and 3° for the back and forth diffuse wave (BF diffuse). The shaded domains on figures \ref{fig:Slab_h2o_h} to \ref{fig:Gran_water_porosity} represent a threshold characterizing a lack of statistic due to the Monte Carlo nature of the model. Its value has been arbitrarily set to 3 rays, meaning features including less rays than the threshold don't have enough statistics to be interpreted. Specular rays are not concerned by this threshold. Dashed lines indicate the time of the first possible back and forth for each case. The bottom limit condition describing the underlying medium has been set to $A=0.15$ \cite{DENEVI_2008,Domingue_2011} to describe average material relevant for Mercury's surface. Moreover, WARPE simulates the travel of the photons inside the surface and the times when each one leaves it, but the travel between BELA and the surface is considered as propagation in vacuum and therefore not modeled. Thus, in WARPE the first specular reflection at the surface happens at time $0\,\mathrm{ns}$. In the figures \ref{fig:Slab_h2o_h} to \ref{fig:Slab_water_S_porosity}, this time 0 has been arbitrarily set to $1.10^{-6}\,\mathrm{ns}$ for readability, as the plots are in log-scale, and not photon that actually gets inside the surface and leaves it before this time.


\subsection{Compact slab texture}\label{subsec:Slab_texture}

This section focuses on compact slab texture, meaning that the surface is made of a material that is filled with matter, with near-spherical impurities (such as void or another material composition). Here the roughness is set to 0.1° following section \ref{subsec:Numerical-experiments}.

\begin{figure}
\centering
\centerline{\hfill{}\includegraphics[scale=0.5]{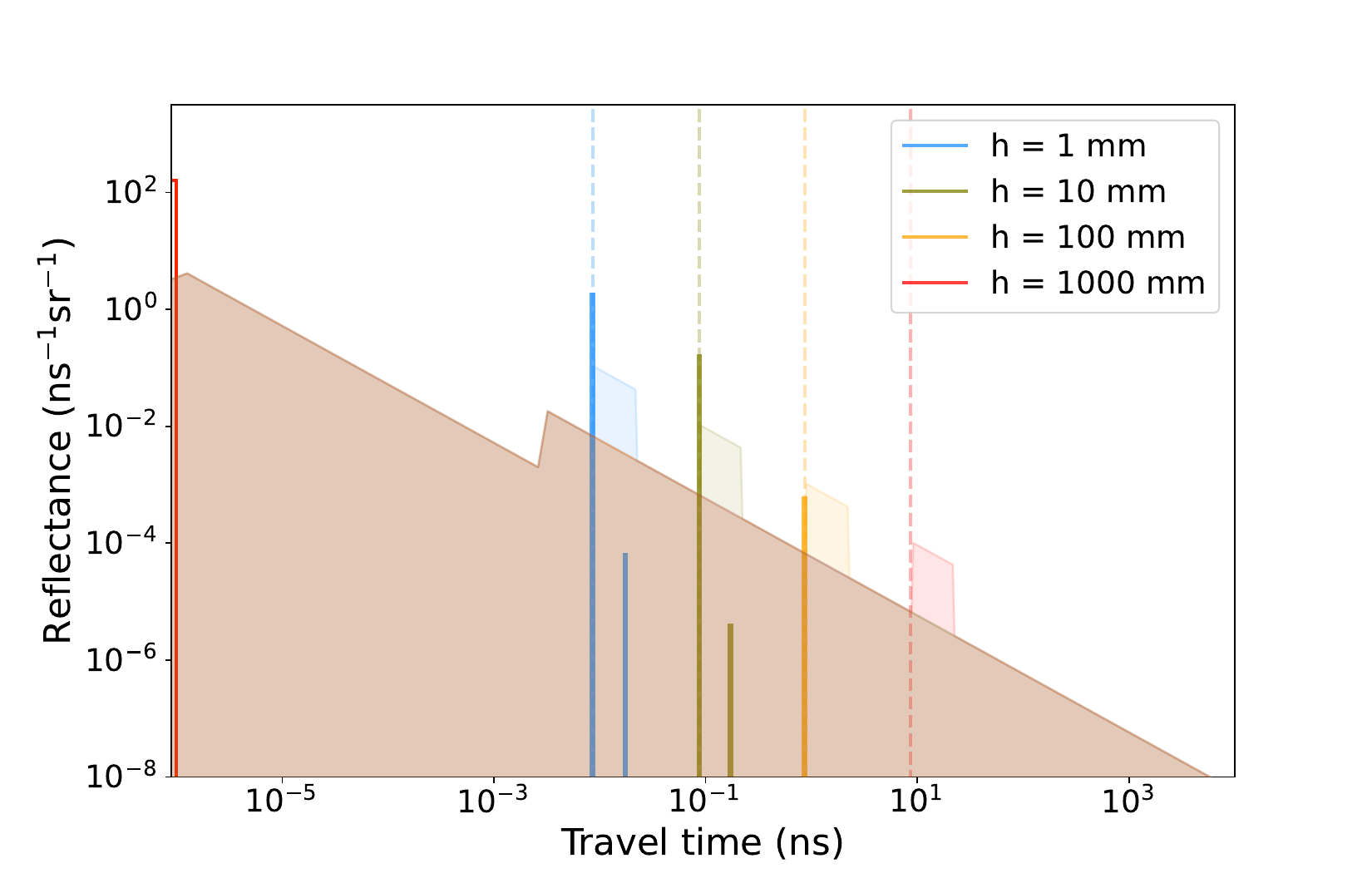}\hfill{}}
\caption{Effect of physical thickness on the waveform in pure water ice compact slab. The absence of features for $h=1000$ $\mathrm{mm}$ is explained by the medium being optically thick ($\tau=15$) with a significant extinction coefficient $k$ of $1.272\times10^{-6}$. The analytical solution gives a reflectance of $1.47\times10^{-15}$$\mathrm{sr}^{-1}$. The transparent color-filled domain corresponds to the limits of non-constrained waveform due to the limitation of the Monte-Carlo approach of WARPE (see \cite{Barron_2025} for more details). Each color plot has its own domain. When there is a superposition of the domains, the resulting color depends on each individual the color plots and is thus non informative by itself.}
\label{fig:Slab_h2o_h}
\end{figure}

\begin{figure}
\centering
\centerline{\hfill{}\includegraphics[scale=0.5]{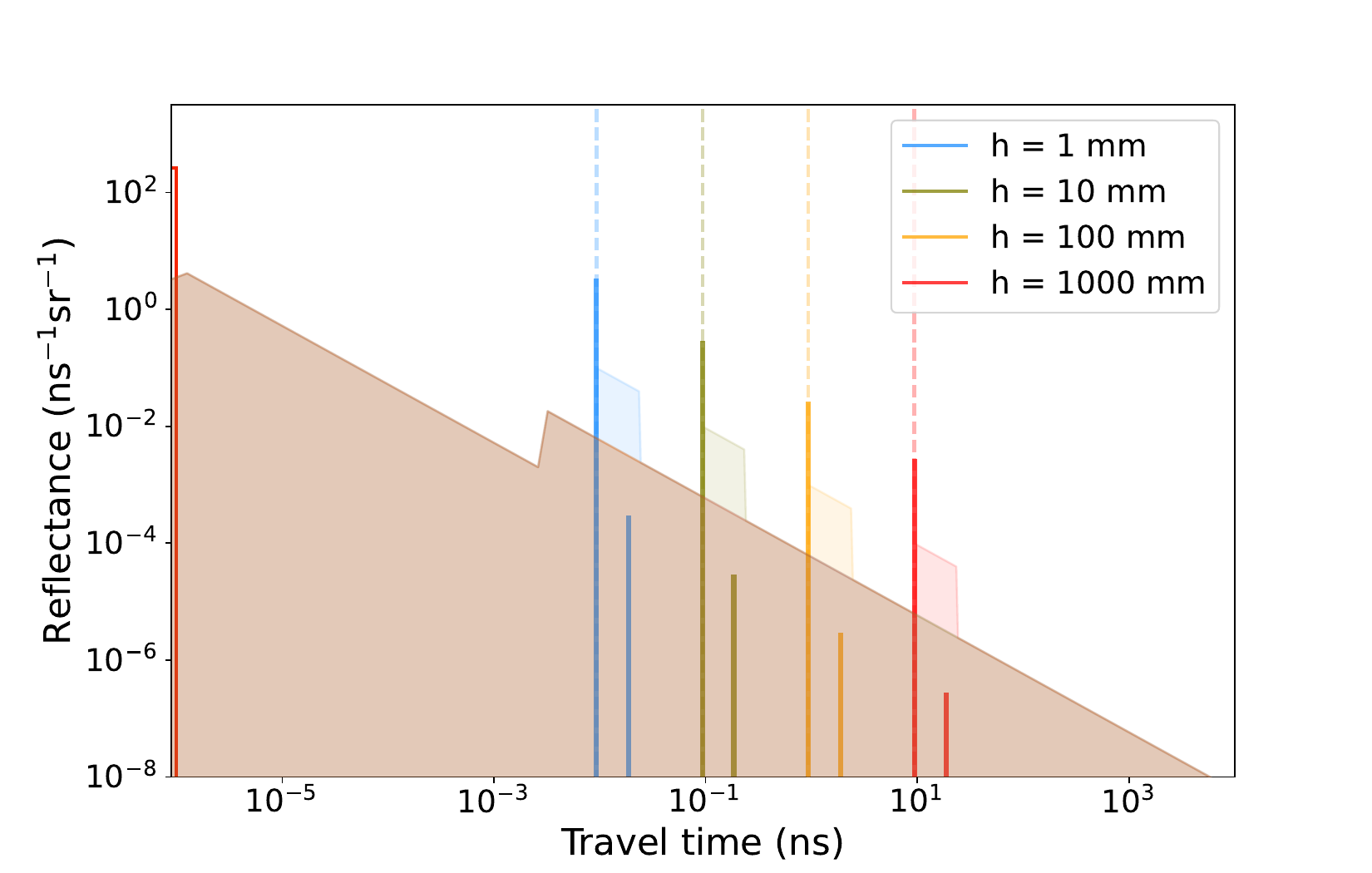}\hfill{}}
\caption{Effect of physical thickness on the waveform in pure CO$_2$ ice compact slab. This time the feature is visible for $h=1000$ $\mathrm{mm}$. The medium is less optically thick ($\tau=0.01$) with an extinction coefficient $k$ of $8.47\times10^{-10}$. The analytical solution gives this time a reflectance value of $0.026$ $\mathrm{sr}^{-1}$. On the logarithmic scale, the return time of the first BF almost reach BELA sampling resolution but in this case the required physical thickness is 1335 $\mathrm{mm}$.}\label{fig:Slab_co2_h}
\end{figure}

For pure slab without impurities, there is no diffusion and the value of the single scattering albedo $\omega$ is 0, therefore no background scattering is observed for this situation whatever the composition as seen in figure \ref{fig:Slab_h2o_h} for water ice and figure \ref{fig:Slab_co2_h} for CO$_2$ ice. In both cases an increasing thickness $h$ results in a larger attenuation of the return pulse. If the material is significantly absorbant and physically thick (for example $k_{H_2O}\sim10^{-6}$ with $h=1000$ m) the medium will completely absorb the light and no BF will leave at the top. 

\begin{figure}
\centering
\centerline{\hfill{}\includegraphics[scale=0.5]{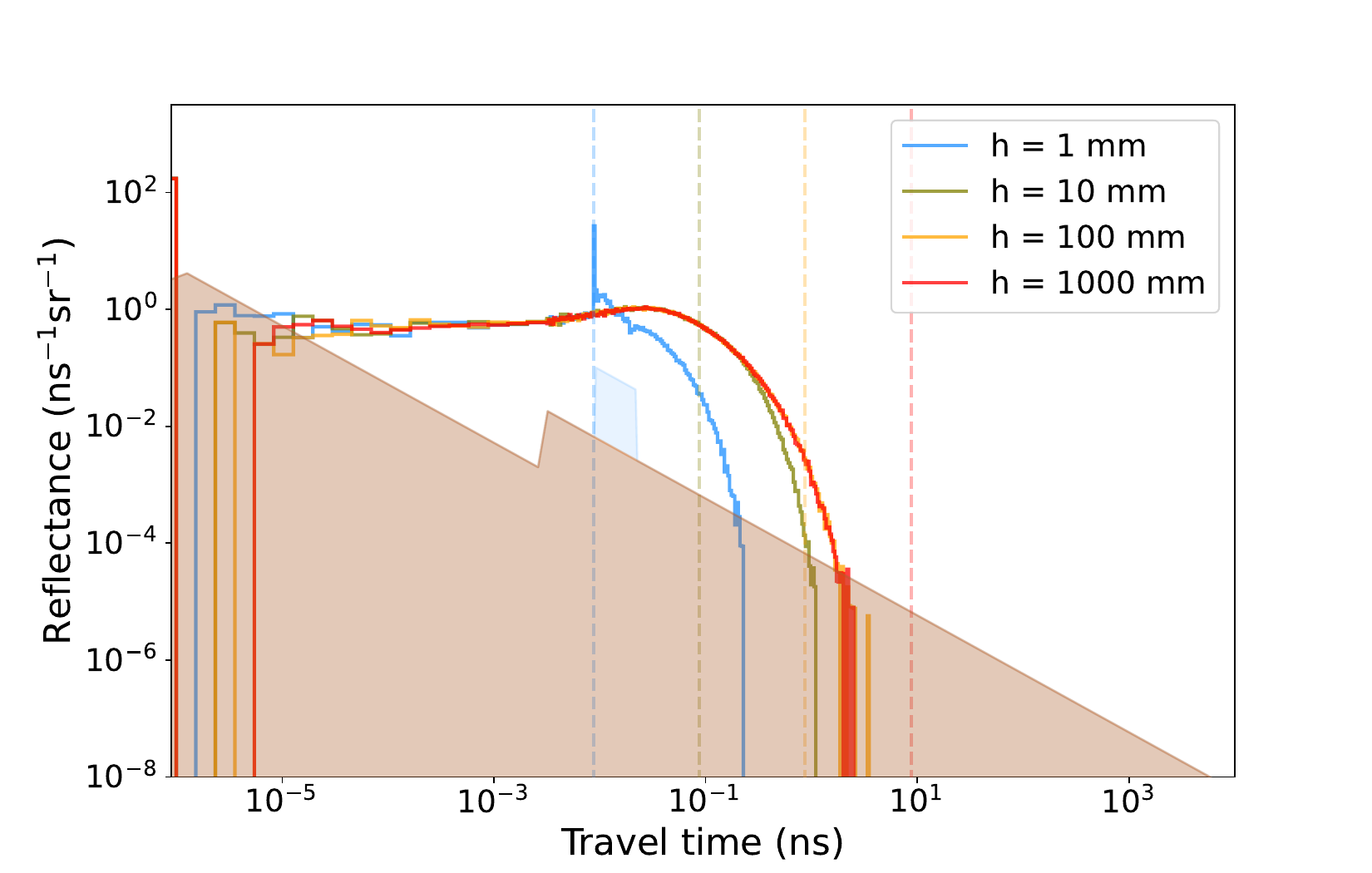}\hfill{}}
\caption{Effect of physical thickness on the waveform in water ice compact slab with $1-\gamma_{c}=$10\% of CO$_2$ impurities of 100 $\mathrm{\mu m}$. The impurities, even less absorbing, have a significant effect on the scattering inside the medium and will therefore smooth the waveform. Comparing to figure
\ref{fig:Slab_h2o_h}, only at $h=1$ $\mathrm{mm}$, an attenuated BF is visible.}
\label{fig:Slab_eau_co2_10_h}
\end{figure}

\begin{figure}[p]
\centering
\centerline{\hfill{}\includegraphics[scale=0.5]{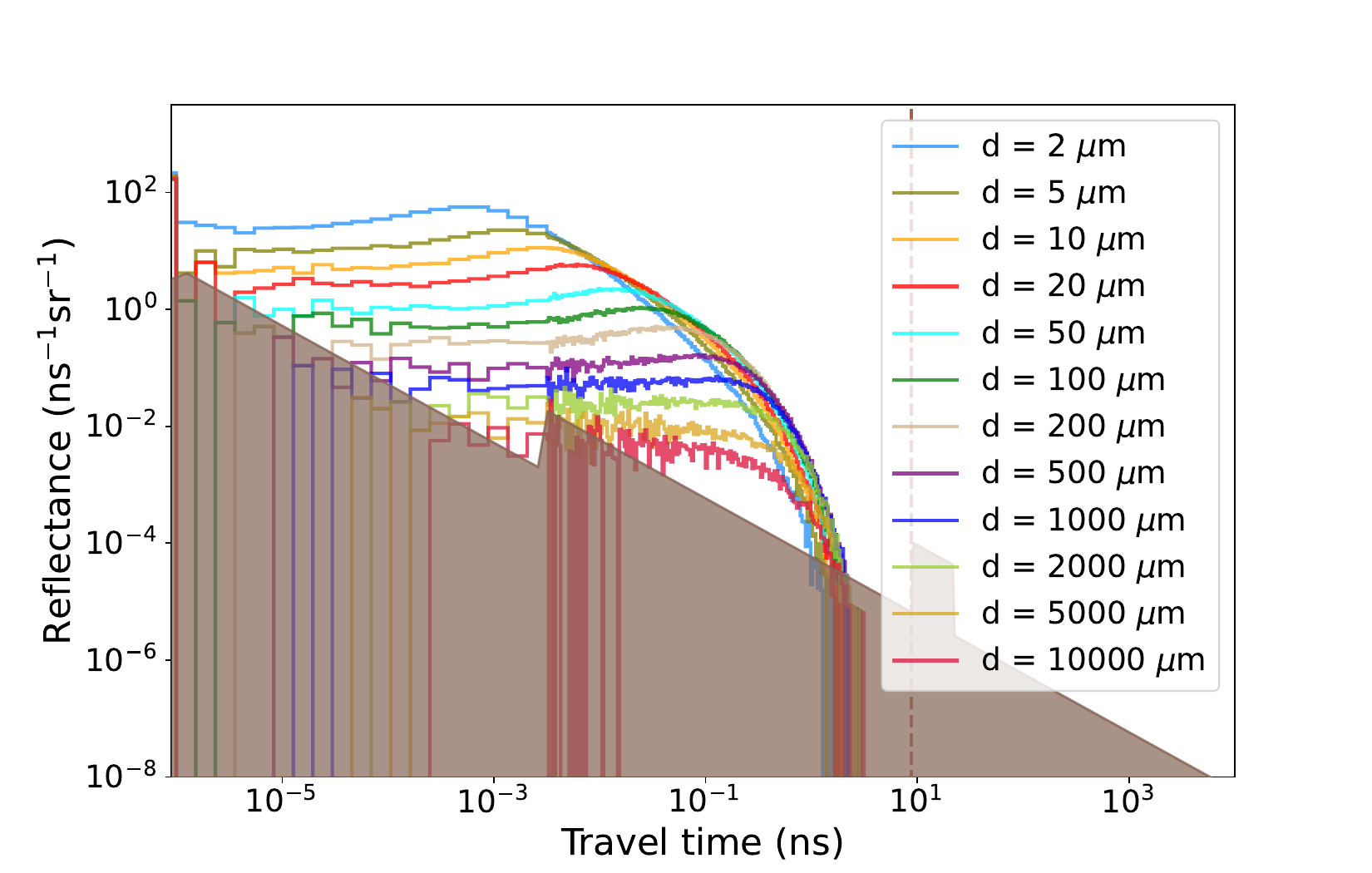}\hfill{}}

\caption{Effect of grain size $d$ of CO$_2$ impurities on the waveform in water ice compact slab with $1-\gamma_{c}=$10\% volumetric impurities and a thickness $h=1000$ mm. As $d$ increases, the absorption becomes more dominant (low value of $\omega$) and the waveform tends to the case without impurities (see Fig. \ref{fig:Slab_h2o_h}, case $h=1000$ mm). When $d$ is small, more scattering occurs then more rays leave the medium and the waveform tends to be shorter. This situation can be considered as optically thick because no rays reach the bottom interface.}
\label{fig:Slab_eau_co2_10_diam}
\end{figure}

Adding impurities in a compact medium introduces background scattering in the return pulse shape since the single scattering albedo will increase. Figure \ref{fig:Slab_eau_co2_10_h} shows that the thickness limit to see a BF in lowered from 100 $\mathrm{mm}$ to 1 $\mathrm{mm}$ when adding impurities of 100 $\mathrm{\mu m}$ diameter. The grain size also affects the diffusion and thus the waveform. Figure \ref{fig:Slab_eau_co2_10_diam} illustrates the role of the grain size, showing that smaller impurities contribute to more scattering. Larger impurities increase the total absorption, simply because the rays cannot escape at the top anymore due to the lack of diffusion.

\begin{figure}[p]
\centering
\centerline{\hfill{}\includegraphics[scale=0.5]{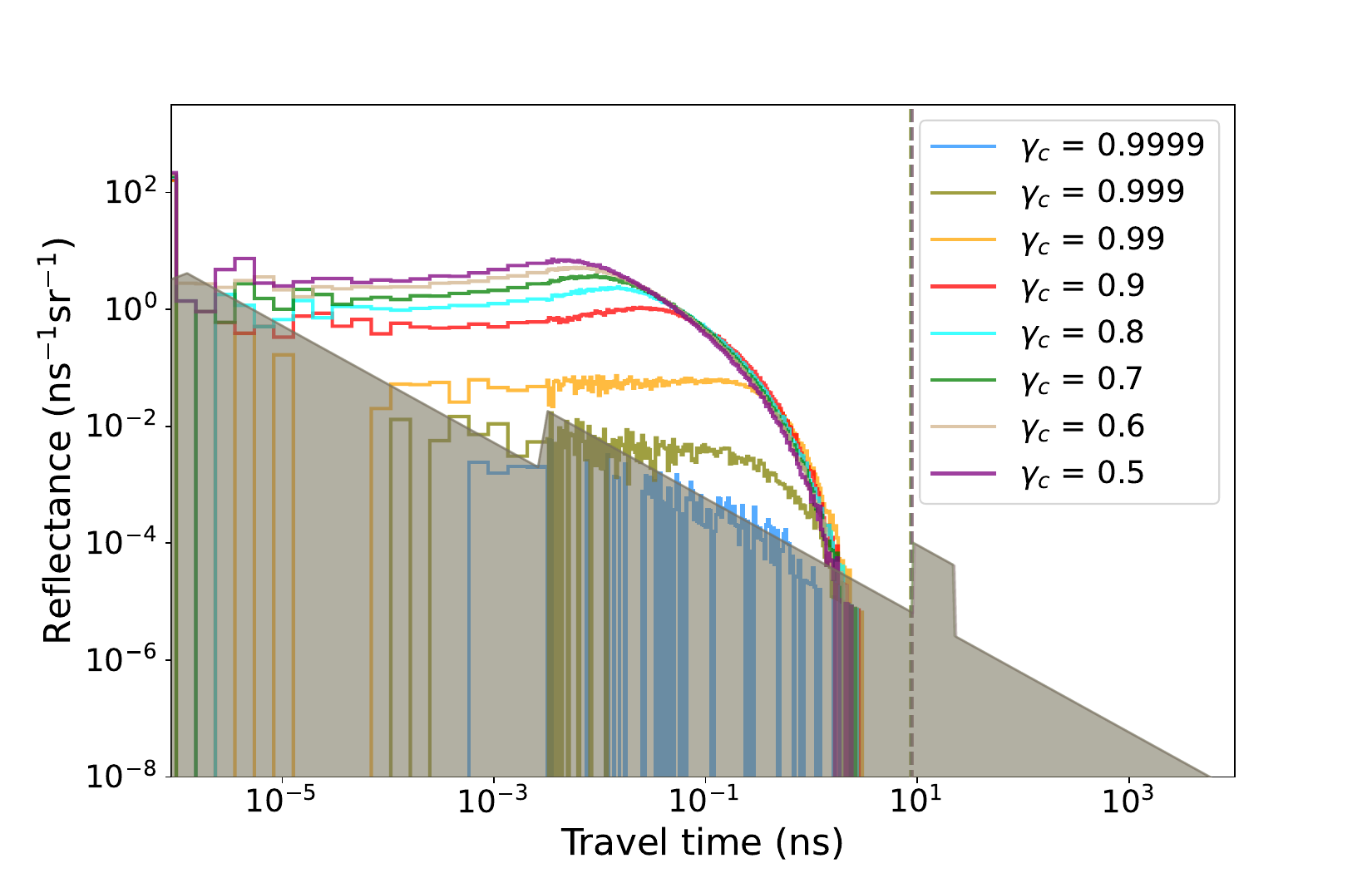}\hfill{}}
\caption{Effect of compacity $\gamma_{c}$ on the waveform in water ice compact slab with CO$_2$ impurities of 100 $\mathrm{\mu m}$ diameter grain size. The medium is optically thick, only the background scattering can be observed. At the extreme case $\gamma_{c}=0.9999$, $\omega$ tends to 0, approaching the situation described in figure \ref{fig:Slab_h2o_h} for $h=1000$ mm. The proportion of impurities slightly changes the value of the effective refractive index $n$ and therefore the efficiency of the refraction and the reflections at interfaces. At 1064 $\mathrm{nm}$, $n_{H_2O}$ < $n_{CO_2}$ and thus more impurities increases the equivalent $n_{eff}$ of the medium for this case.}
\label{fig:Slab_water_co2_porosity}
\end{figure}

\begin{figure}[p]
\centering
\centerline{\hfill{}\includegraphics[scale=0.5]{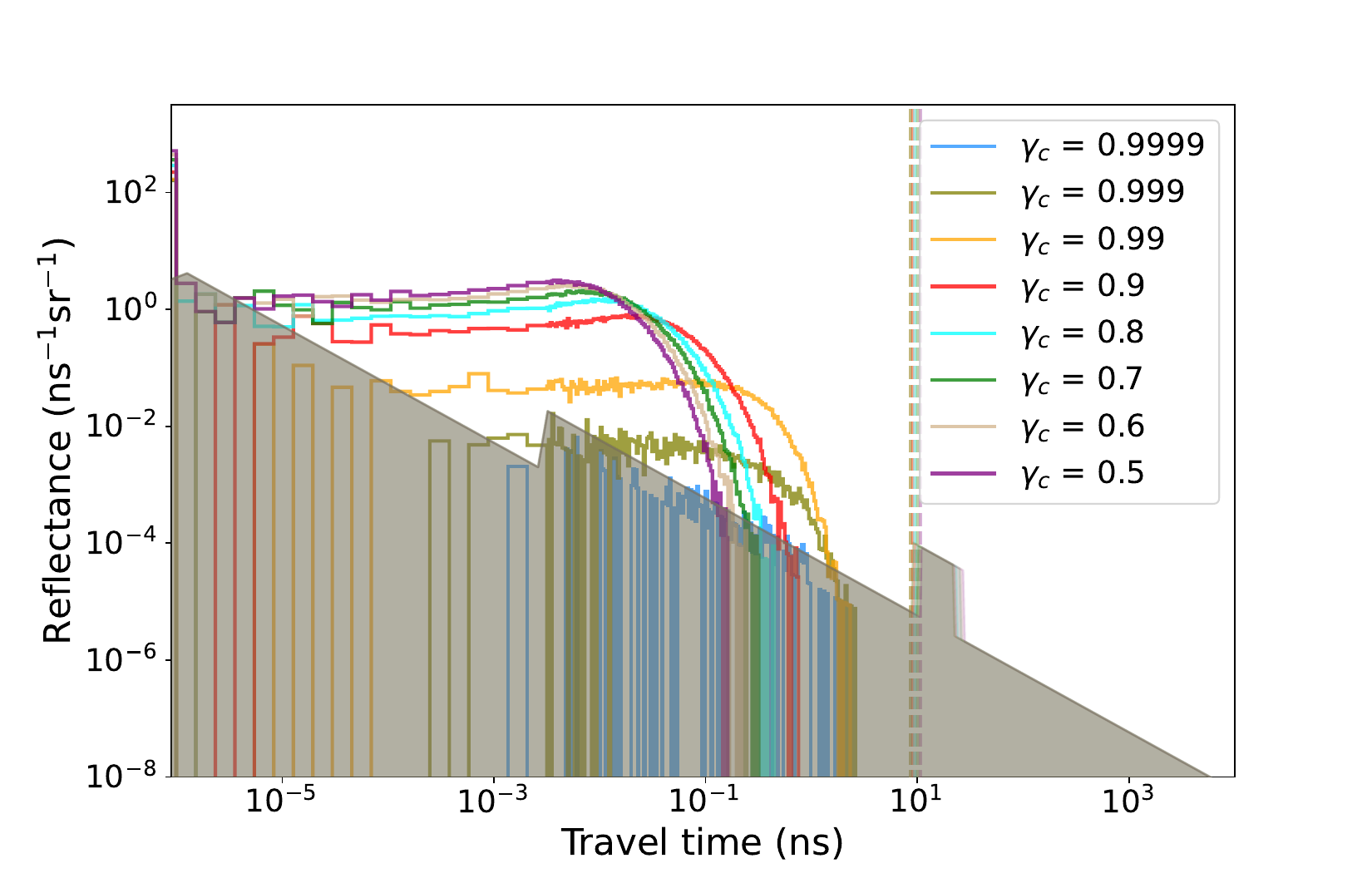}\hfill{}}
\caption{Effect of compacity $\gamma_{c}$  on the waveform in water ice compact slab with S impurities of 100 $\mathrm{\mu m}$ diameter grain size. The medium is also optically thick, only the background scattering can be observed. The more impurities (smaller $\gamma_{c}$), the more scattering, the larger the reflectance and the shorter mean travel time.  This effect is highly non-linear as expected. At the extreme case $\gamma_{c}=0.9999$, $\omega$ tends to 0, approaching the situation described in figure \ref{fig:Slab_h2o_h} for $h=1000$ mm. By comparing with Fig. \ref{fig:Slab_water_co2_porosity}, the contrast of $n$ being more important between Sulfur and H$_2$O the scattering is more efficient, travel-time in the medium is shortened.}
\label{fig:Slab_water_S_porosity}
\end{figure}

\begin{figure}[p]
\centering
\centerline{\hfill{}\includegraphics[scale=0.5]{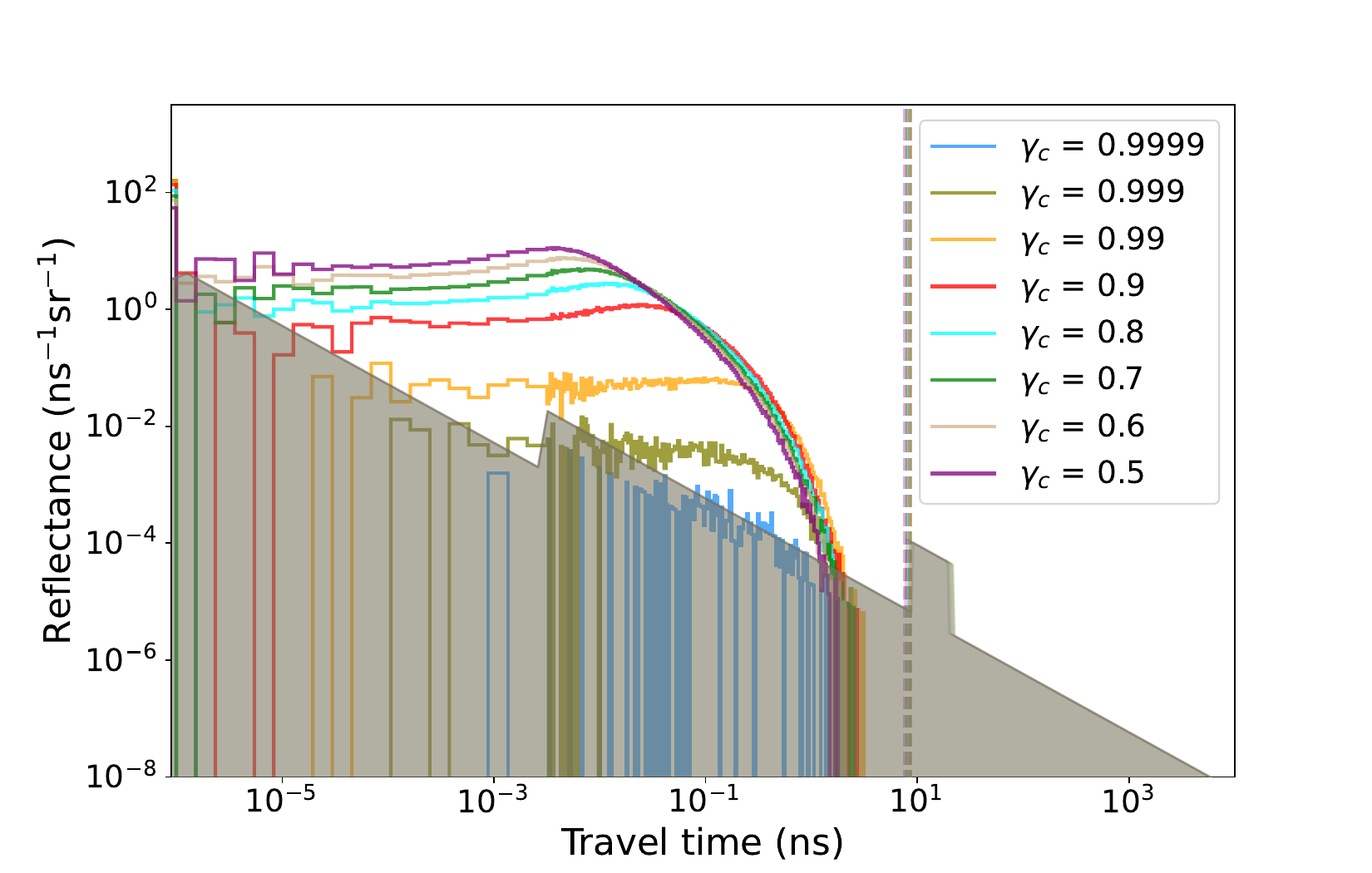}\hfill{}}
\caption{Effect of compacity $\gamma_{c}$  on the waveform in water ice compact slab with void impurities of 100 $\mathrm{\mu m}$ diameter grain size. The medium is optically thick too, therefore apart from the the specular reflection, only the background scattering can be observed. Void impurities decrease the intensity of the specular reflection as its refractive index is lower than the slab's. Moreover, low absorbent material and void tend to have similar properties on the pulse shape since diffusion will be more frequent than absorption according to the value of the extinction coefficient.}
\label{fig:Slab_water_void_porosity}
\end{figure}

The compacity $\gamma_{c}$ (or filling factor) of the compact slab media also largely influence the waveform (see figure \ref{fig:Slab_water_co2_porosity} and \ref{fig:Slab_water_S_porosity}.) The more impurities (smaller $\gamma_{c}$), the more scattering, the larger the reflectance and the shorter mean travel time, as expected. This effect is highly non-linear, with more difference at high compacity (low 1-$\gamma_{c}$ volumetric impurities content). The effect of the impurity composition (optical indexes) can be noted by comparing Fig \ref{fig:Slab_water_co2_porosity}, \ref{fig:Slab_water_S_porosity} and \ref{fig:Slab_water_void_porosity}. Sulfur is more absorbing than CO$_2$ and thus the waveform reflectance is thus lower and the mean travel time is shorten.

\newpage
\subsection{Granular texture}\label{subsec:Granular_texture}

This section focuses on granular texture, i.e. near spherical grains filling the volume. The grains can be of different compositions and the porous space is vacuum, contrarily to the previous compact slab case. This type of configuration doesn't include any interface at the top therefore only background scattering is expected and no specular reflection, nor BF direct.

\begin{figure}[h!]
\centering
\centerline{\hfill{}\includegraphics[scale=0.5]{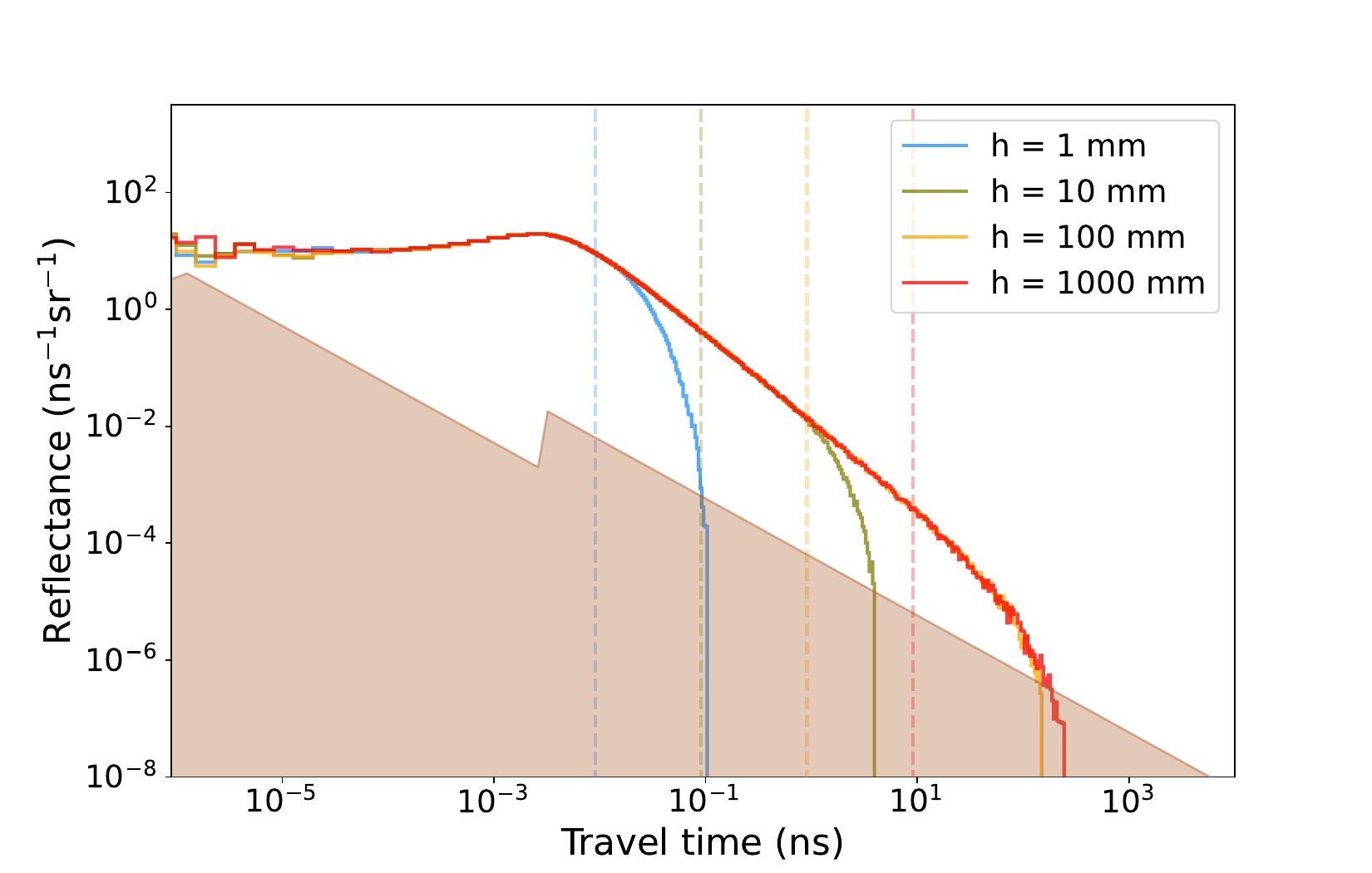}\hfill{}}
\caption{Effect of physical thickness $h$ on the waveform in pure CO$_2$ ice granular medium with a grain size of 200 $\mathrm{\mu m}$ and compacity of $\gamma_{c}=0.9$. CO$_2$ ice at 1064 $\mathrm{nm}$ is very low absorbing medium ($k_{CO_2}\sim10^{-6}$) but the bottom layer is very absorbing ($A=0.15$). Interestingly, the waveform distribution is not changed for thicker $h$ in the shorter time domain, but rather after a threshold time that depends on $h$. When the medium becomes optically thick, ($h\gtrsim 100$ mm) the waveform is not dependent on $h$ anymore.}
\label{fig:Gran_co2_h}
\end{figure}

The thickness $h$ of granular medium, on top of a relatively dark substratum $A=0.15$ is studied in Figure \ref{fig:Gran_co2_h}. It indicates that for a very low absorbing medium ($k_{CO_2}\sim10^{-10}$), the travel time distribution is almost not changed with $h$ but simply extended in the largest time. Since the substratum is significantly absorbing, if a ray reach the substratum, it has high chances of being absorbed. Increasing $h$ lead thus to less bottom absorption. This effect saturates when the medium is optically thick, in this case $h=100$ mm for CO$_2$ grains at 200 $\mathrm{\mu m}$ diameter and compacity of $\gamma_{c}=0.9$. 

\begin{figure}[h!]
\centering
\centerline{\hfill{}\includegraphics[scale=0.5]{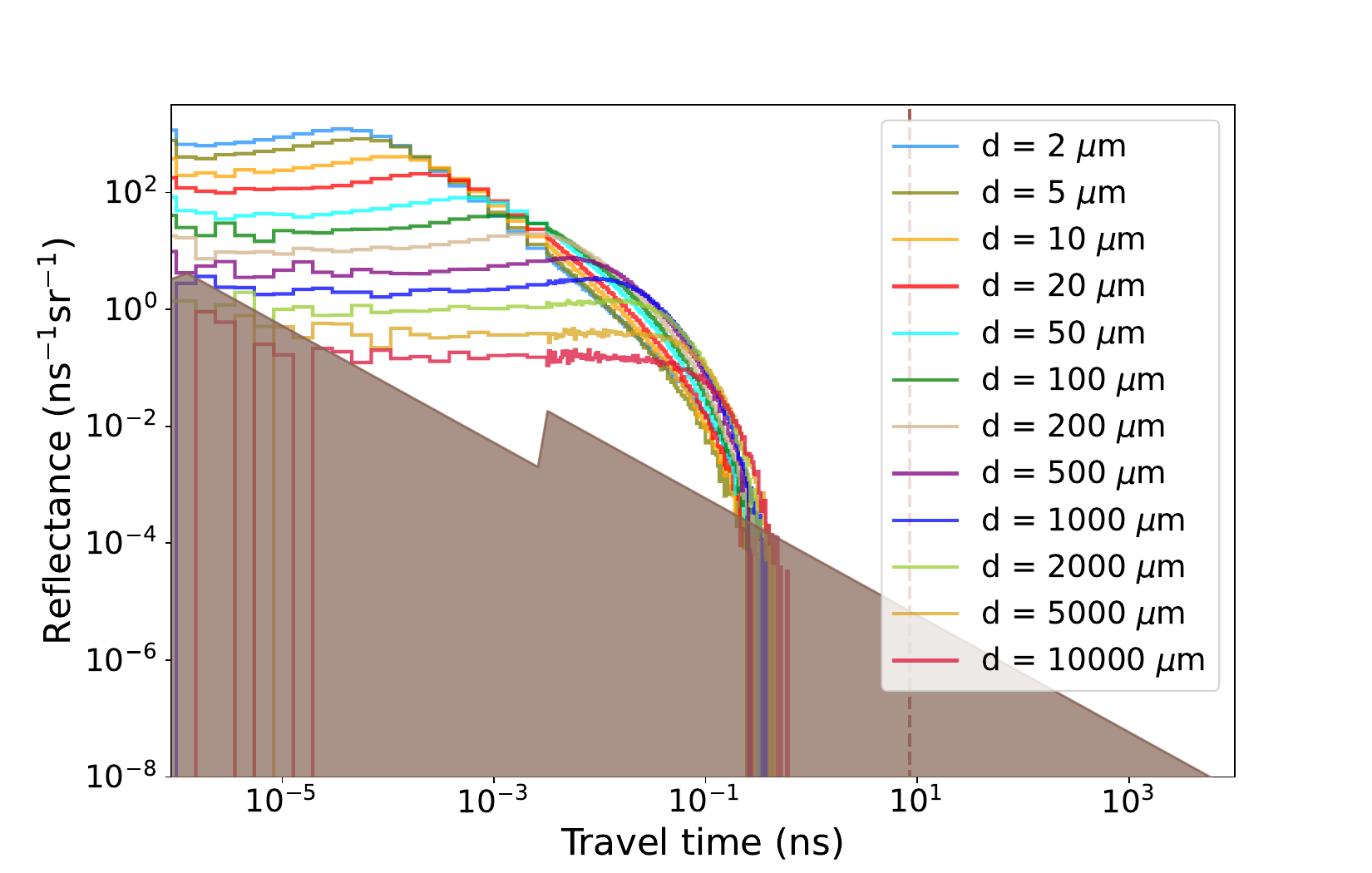}\hfill{}}
\caption{Effect of grain size $d$ on the waveform in pure water ice granular medium with a thickness $h=$1000 $\mathrm{mm}$ and $\gamma_{c}=0.9$. The same effect is observed as in figure \ref{fig:Slab_eau_co2_10_diam} for compact slab ice, except here without interface: rays tend to leave sooner the medium and the reflectance (in $\mathrm{ns}^{-1}$$\mathrm{sr}^{-1}$) is more intense with smaller grain size $d$, due to more scattering.}
\label{fig:Gran_eau_diam} 
\end{figure}

The effect of grain size (similar to Fig. \ref{fig:Slab_eau_co2_10_diam} for compact slab ice) is studied in Figure \ref{fig:Gran_eau_diam}. At a constant compacity $\gamma_{c}$, the medium contains more grains when grain size is small and therefore scattering occurs more often. This effect results in more rays leaving sooner the medium with more intense reflectance. 

\begin{figure}[h!]
\centering
\centerline{\hfill{}\includegraphics[scale=0.5]{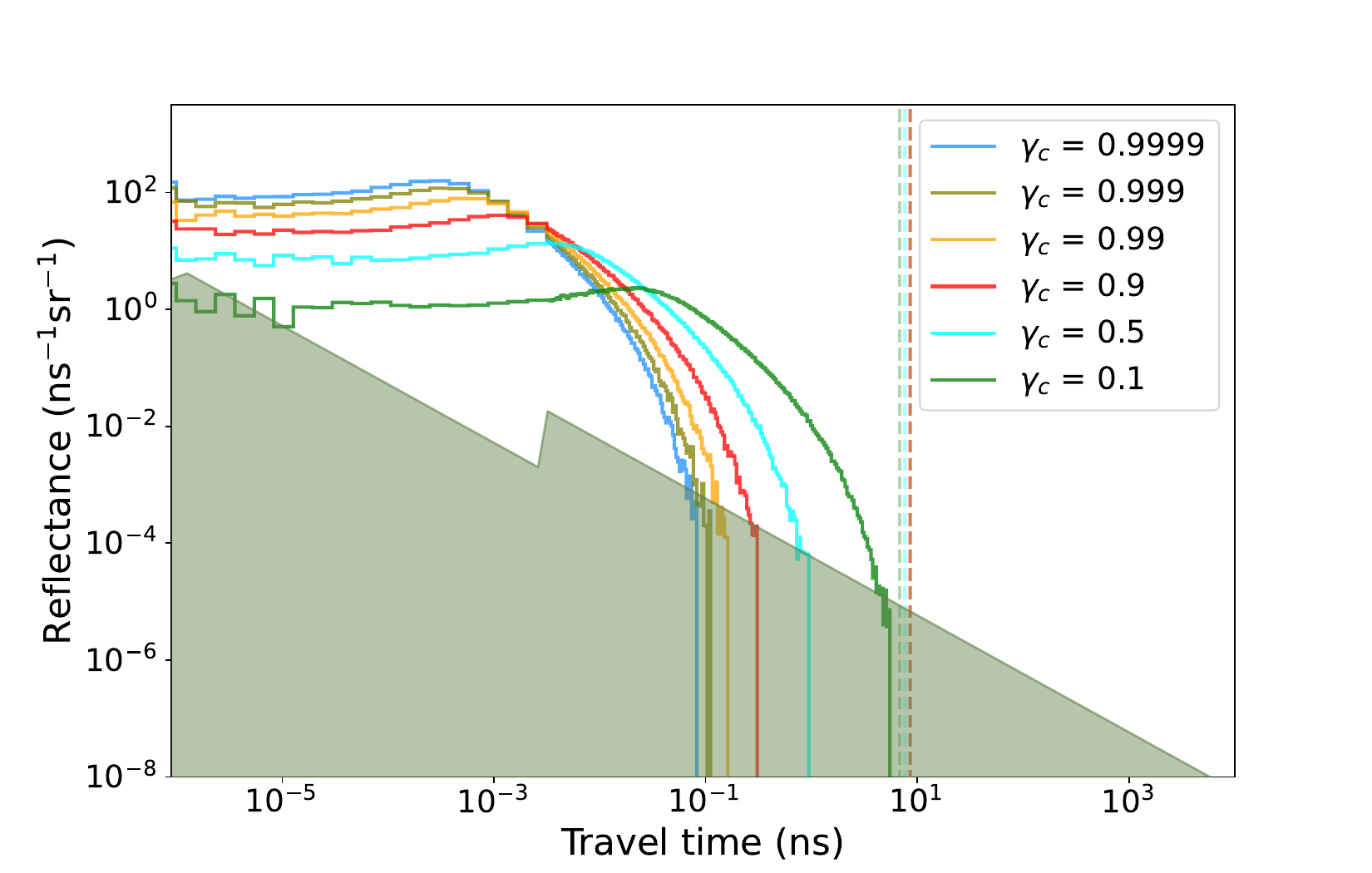}\hfill{}}
\caption{Effect of compacity on the waveform in pure granular water ice with a thickness of 1000 $\mathrm{mm}$ and a grain size of 100 $\mathrm{\mu m}$. A compacity of $\gamma_{c}=0.1$ means the medium is very porous. Rays will encounter less grain and the time residence will then increase.
}
\label{fig:Gran_water_porosity}
\end{figure}

Figure \ref{fig:Gran_water_porosity} shows the effects of compacity. The compacity $\gamma_{c}$ of a granular medium represents the proportion of materials against voids. For this reason the range of possible $\gamma_{c}$ is larger for granular texture in comparison with compact texture. First, the larger the compacity, the larger the scattering simply because there are more grains in the same volume. This effect produces shorter and more intense waveform for larger compacity $\gamma_{c}$.

\subsection{Effective speed of light in complex media}

Effective light velocity alter the waveform by definition. Indeed complex material mixture has a varying effective speed of light which is derived from effective refractive index as seen in section \ref{subsec:Simulation}. For granular material, the higher the compacity means less void, thus slower effective speed of light. Adding other types of impurities in the mixture affect the effective speed of light depending on end-member's proportions and values of refractive index. For compact slab texture, the effect is similar and also depends on the impurities composition. If the light is faster in impurities than the slab ($n_{imp}<n_{slab}$), $n_{eff}$ is lower and conversely when $n_{imp}>n_{slab}$. From  Figure \ref{fig:Slab_water_co2_porosity}, \ref{fig:Slab_water_S_porosity}, \ref{fig:Slab_water_void_porosity} and \ref{fig:Gran_water_porosity}, it seems that this effect is not dominant. For instance Figure \ref{fig:Gran_water_porosity} shows shorter travel time for larger compacities. This shows that the scattering effect is dominant over this effective speed of light effect.

\subsection{BELA measurements}
\label{sec:BELA_measurement}

The time resolution (sampling interval) of BELA is $12.5\,\mathrm{ns}$ \cite{Thomas_2021}. This implies that the shape of any feature that doesn't expand longer than this resolution cannot be resolved, as all the signal would be condensed in one sampling step. According to previous section \ref{subsec:Slab_texture} and \ref{subsec:Granular_texture}, the only case that allow a BELA observation (with significant reflectance waveform beyond $12.5\,\mathrm{ns}$) are those including only low absorbing material at 1064 $\mathrm{nm}$ such as CO$_2$. If present in Mercury's conditions, CO$_2$ may appear only in the coldest region since its sublimation pressure is low \cite{Fray_2009}, and implies a low sublimation rate \cite{Zhang_2009,Zhang_2010}. Nevertheless, any volatile compound with such low absorption properties would optically behave in an equivalent manner. The objectives here is to understand how instrument data acquisition affects WARPE simulation and also to show how compact slab and granular medium can be distinguished using BELA. 

\begin{figure}
\centering
\centerline{\hfill{}\includegraphics[scale=0.5]{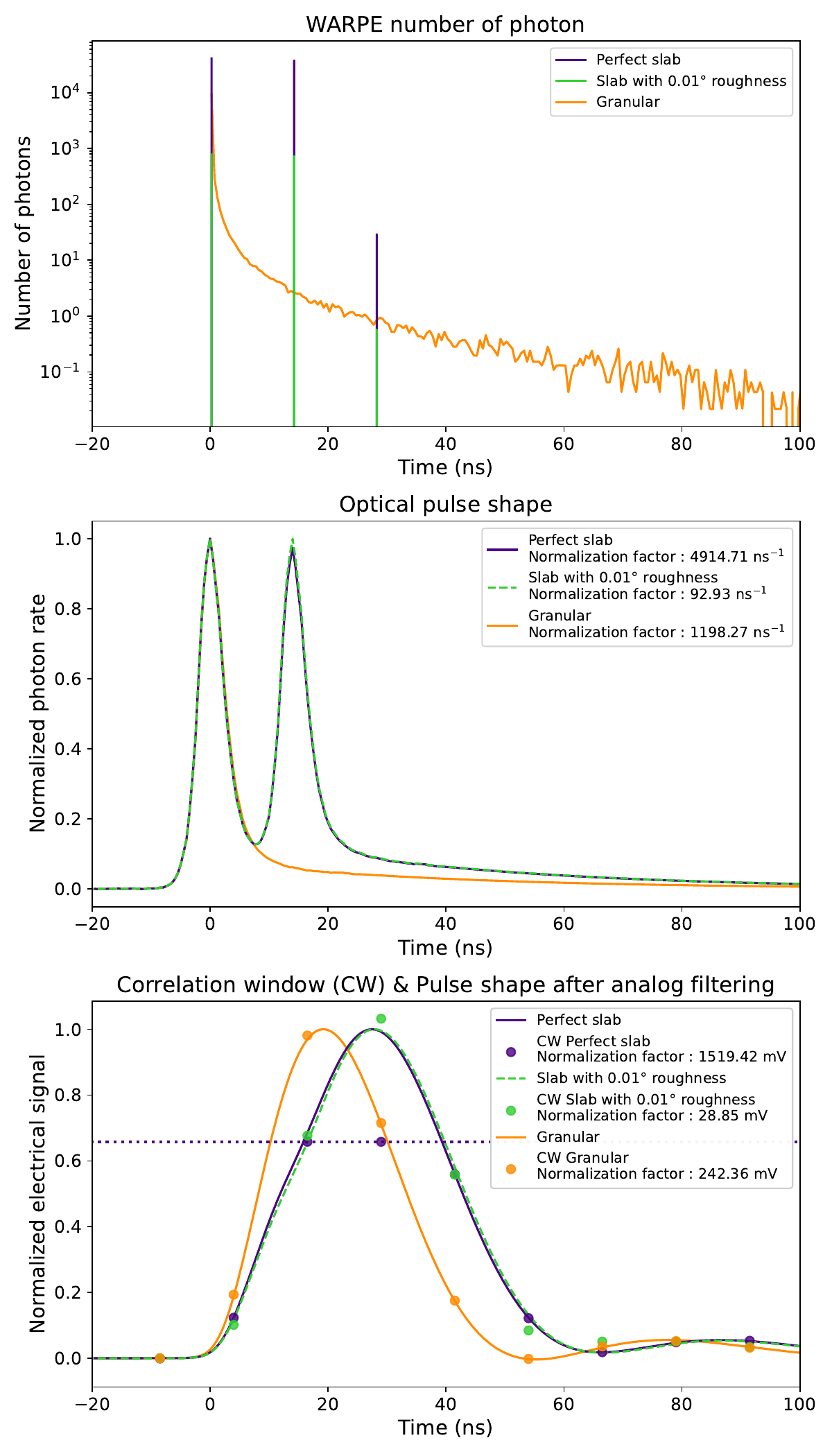}\hfill{}}
\caption{Realistic BELA waveform using instrumental modeling and WARPE simulations. (top) WARPE simulation of a perfect instrument with 3 cases: 1000 $\mathrm{mm}$ granular medium of pure CO$_2$ ice with grain size of 200 $\mathrm{\mu m}$ and $\gamma_{c}=0.9$ noted Granular (similar to Fig. \ref{fig:Gran_co2_h}) ; 1500 $\mathrm{mm}$ pure slab of CO$_2$ with no roughness noted perfect slab and finally the same slab with a roughness of 0.01° (similar to Fig. \ref{fig:Slab_co2_h}). This graph represented in log scale.
(middle) The photons rate in $\mathrm{ns}^{-1}$$\mathrm{sr}^{-1}$ reflected from the surface using realistic outgoing pulse shape from BELA (perfect receiver) using BELA simulation chain \cite{Nishiyama_PulseShapeRoughness_2026}. Granular medium present a single peak, whereas the slab medium present 2 peaks due to the specular reflection at the top and at the bottom (BF direct). The third peak is negligible. Interestingly, the first peak shape is very similar in the slab and granular case.
(bottom) The electric signal shape after analog filtering expected including both transmitter and receptor specification using BELA simulation chain \cite{Nishiyama_PulseShapeRoughness_2026}. This signal sampled at 12.5 ns is the one recorded by the BELA instrument. Granular and slab signals do not reach their maximum at the same time and their shape significantly differ due to the presence/absence of a second peak in optical pulse shape. This difference should be exploited to estimate the mictrotexture of the icy surface. The shape of the signal for slab with different roughness is similar however their intensities varies (Normalization factor are displayed). The purple dotted line indicates saturation threshold of 1000 mV from BELA, that is reached by perfect slab case. Roughness larger than 0.1° strongly attenuates the signal and cannot be recorded by BELA (not displayed here). The green (and orange to a lower extent) dots do not perfectly match the curves due to added random noise in the signal sampled by BELA.}
\label{fig:WARPE_BELA}
\end{figure}

Figure \ref{fig:WARPE_BELA} illustrates WARPE results used to run the BELA simulation chain proposed and detailed in \cite{Steinbruegge2018,HosseiniArani_Comprehensiveorbitperformance_PaSS2021, Nishiyama_PulseShapeRoughness_2026}. The result show that slab and granular can be distinguished in the final recorded electric signal. In addition, the specular reflection is mainly controlled by the surface roughness that spread out the specular lobe. We demonstrate that a roughness larger than 0.1° ``dilutes'' heavily the specular lobe so that not enough photons can reach the BELA instrument anymore. This lack of signal can thus is indirectly identified as the presence of a rough slab. Another result is that a $1500\,\mathrm{mm}$ thick low absorptive slab is enough to shift the maximum of the instrumental response by one sample step. Interestingly, it suggests that thinner slabs could be identified by their way of delaying the maximum of the measured electrical signal.

\newpage
\section{Discussions and conclusion}

We studied for the first time the fundamental time response of the optical pulse shape as a function of the microtexture. We show how physical properties influence the waveform of the return pulse shape. Simulations reveal the effect of grain size, compacity and thickness. In the case of complex material texture, we demonstrate that the scattering effect is dominant over the effective speed of light.

Specular reflection can be observed only for compact slab in various microtexture and composition with its intensity being highly roughness-dependent. According to the simulations granular and slab configuration cannot be distinguished. The presence of the specular lobe is the only way to determine whether the medium is granular or a slab. If the roughness is too important, the intensity of the specular reflection drops. So on the first order, roughness is the predominant parameter in microtexture regarding the reflectance.

We propose for the first time to assess quantitatively the waveform expected for different microtexture of volatiles material on Mercury in the scope of BELA \cite{Thomas2007} onboard BepiColombo. By adapting the WARPE simulation \cite{Barron_2025} on compact and granular texture, we demonstrate that only very low absorbing material, such as CO$_2$ ($k_{CO_2}\sim10^{-10}$) are able to be actually deciphered in the time domain. Water ice ($k_{H_{2}O}\sim10^{-6}$) is too absorbent at 1064 $\mathrm{nm}$ to get a significant amount of photons coming back from the bottom. 

Real data from BELA, especially on PSR, will be our next analysis target. We envisaged to create a database of scenarii that will be then compared to real data, following the same strategy as for the surface slope distribution \cite{Nishiyama_PulseShapeRoughness_2026}.

This work could also apply on the similar instrument GALA \cite{Hussmann2019, Hussmann_GanymedeLaserAltimeter_SSR2025} onboard JUICE which has a time resolution of 5 ns. This instrument would thus be able to decipher the microtexture in a broader range of parameters. Similar work could also be achieved on Earth's ice caps with ATLAS instrument on board of the ICESat-2 mission \cite{Abdalati2010}. These icy surfaces differ from PSR as they tend to be most likely constituted of a slab with impurities rather than icy regolith. 

Our work is limited by the availability of optical constants of material. Here we only tested water ice, CO$_2$ and Sulfur due to the lack of properties available in the literature. This work clearly encourage a laboratory work to produce high quality optical constants for planetary science relevant materials.

In a future, we emphasize that a next generation of spaceborne full-waveform laser instrument with nanosecond time resolution or lower would contribute to give more detailed insights on the microphysics of planetary surfaces. Another potential improvement could also to select and optimize the laser wavelength (or even multi-wavelength) in order to follow a particular material of interest.

\section*{Declarations}

\section*{Availability of data and materials}
All simulation data may be shared upon reasonable requests to corresponding author (JB).

\section*{Competing interests}
The authors declare that they have no competing interests.

\section*{Funding}
We acknowledge support from the ``Institut National des Sciences de l'Univers'' (INSU), the ``Centre National de la Recherche Scientifique'' (CNRS) and ``Centre National d'Etudes Spatiales'' (CNES) through the ``Programme National de Plan{\'e}tologie''. This work was supported by the Île-de-France Region with DIM origines.
This work has received support from France 2030 through the project
named Académie Spatiale d'Île-de-France (\url{https://academiespatiale.fr/})
managed by the National Research Agency under bearing the reference
ANR-23-CMAS-0041.

\section*{Author contributions}
JB, FS and FA developed methodology and codes for numerical simulations. GN performed simulations including instrumental effects. JB conducted all the numerical simulations, wrote the original draft, and finalized the initial version of the manuscript manuscript. All the authors reviewed, approved the manuscript and contributed to the discussions.

\section*{Acknowledgment}

We acknowledge support from the ``Institut National des Sciences de l'Univers'' (INSU), the ``Centre National de la Recherche Scientifique'' (CNRS) and ``Centre National d'Etudes Spatiales'' (CNES) through the ``Programme National de Plan{\'e}tologie''. This work was supported by the Île-de-France Region with DIM origines.
This work has received support from France 2030 through the project
named Académie Spatiale d'Île-de-France (\url{https://academiespatiale.fr/})
managed by the National Research Agency under bearing the reference
ANR-23-CMAS-0041.

\pagebreak{}

\bibliographystyle{elsarticle-num}
\phantomsection\addcontentsline{toc}{section}{\refname}\bibliography{biblio}

\end{document}